# Developing a multi-variate prediction model for COVID-19 from crowd-sourced respiratory voice data

Yuyang Yan[1*], Wafaa Aljbawi[1], Sami O. Simons[2], Visara Urovi[1]

[1]Institute of Data Science, Maastricht University, 6229 EN Maastricht, The Netherlands

[2]Department of Respiratory Medicine, Maastricht University Medical Center, Maastricht University, 6229 HX Maastricht, The Netherlands

***Correspondence:** Yuyang Yan, Institute of Data Science, Maastricht University, 6229 EN Maastricht, The Netherlands. yuyang.yan@maastrichtuniversity.nl

## Abstract

**Aim:** COVID-19 has affected more than 223 countries worldwide and in the post-COVID era, there is a pressing need for non-invasive, low-cost, and highly scalable solutions to detect COVID-19. This study focuses on the analysis of voice features and machine learning models in the automatic detection of COVID-19.

**Methods:** We develop a deep learning model to identify COVID-19 from voice recording data. The novelty of this work is in the development of deep learning models for COVID-19 identification from only voice recordings. We use the Cambridge COVID-19 Sound database which contains 893 speech samples, crowd-sourced from 4,352 participants via a COVID-19 Sounds app. Voice features including Mel-spectrograms and Mel-frequency cepstral coefficients (MFCC) and convolutional neural network (CNN) Encoder features are extracted. Based on the voice data, we develop deep learning classification models to detect COVID-19 cases. These models include long short-term memory (LSTM), CNN and Hidden-Unit BERT (HuBERT).

**Results:** We compare their predictive power to baseline machine learning models. HuBERT achieves the highest accuracy of 86% and the highest AUC of 0.93.

**Conclusions:** The results achieved with the proposed models suggest promising results in COVID-19 diagnosis from voice recordings when compared to the results obtained from the state-of-the-art.

## Introduction

The coronavirus or severe acute respiratory syndrome coronavirus 2 (SARS-CoV-2) can cause severe respiratory disease to humans and has become a potential threat to human health and the global economy.

On January 2024, more than 6.9 million deaths from coronavirus disease 2019 (COVID-19) are confirmed [1]. Since the outbreak in late December 2019, multiple variants of the virus have spread worldwide [2].

One of the most concerning aspects of SARS-CoV-2 is its rapid spread, the virus easily spreads through surfaces [3], air, breathing, talking, or physical contact [4], and thus it is possible to affect large populations in a very short time. As a result, it is important to quickly identify who is infected. Several research efforts have been carried out to avoid the rapid spread of the epidemic and effectively control the number of infected people [5, 6]. To facilitate the detection of positive cases, several researchers have been exploring the possibility of utilizing auditory data produced by the human body (such as breathing [7], heart rate [8], and vibration sounds [9]) to diagnose and track disease progression [10].

The most common symptoms include fever, dry cough, loss of smell and taste, headache, muscle aches, diarrhea, conjunctivitis, and in more severe cases, shortness of breath, chest pain, and loss of speech or movement [11]. The respiratory tract is impacted which results in a lack of speech energy and a loss of voice due to shortness of breath and upper airway congestion [12]. Recurrent dry coughs can also cause alterations in the vocal cords, reducing the quality of one's voice. A recent study found that individuals with COVID-19 have changes in their voice's acoustic characteristics due to inadequate airflow through the vocal tract as a result of pulmonary and laryngological dysfunction [12]. Consequently, COVID-19 as a respiratory condition may cause patients' voices to become distinctive, leading to identifiable voice signatures.

With the emergence of COVID-19, several attempts have been made to complement standard testing procedures with effective automated diagnosis solutions. The World Health Organization (WHO) considers nucleic acid-based real-time reverse transcription polymerase chain reaction (RT-PCR) to be the standard technique for diagnosing COVID-19 [11]. Due to the high cost and restricted access to the test, as well as the risk of exposing healthcare professionals and medical staff to the virus, it is also infeasible to go to medical centers and undergo RT-PCR after every cough or uncomfortable feeling [11]. The rapid antigen test (RAT) is an alternative that does not require laboratory processing and eliminates the time constraint of RT-PCR. However, its sensitivity declines with lower viral loads, resulting in false negative results in patients with lower levels of the SARS-CoV-2 virus [13].

In this research, we extract Mel-spectrograms, Mel-frequency cepstral coefficients (MFCC) features and convolutional neural network (CNN) Encoder features from the Cambridge COVID-19 Sound database. We develop deep learning models including long short-term memory (LSTM), CNN, and Hidden-Unit BERT (HuBERT) to detect COVID-19 and compare their prediction performance with baseline models including logistic regression (LR) and support vector machine (SVM). To validate the performance of the proposed models, the Coswara dataset is used for validation. Several works have investigated COVID-19 detection with audio recordings such as [5, 14–17]. The novelty of our work is that we train our models only with speech recordings, without relying on breathing, cough, speech, or any symptoms or hospitalization information. Besides, we develop both deep learning classification models and traditional machine learning models to determine which model is optimal for diagnosing COVID-19. Furthermore, we validate the solution with an external dataset.

The reminder of the paper is organized as follows. Background describes the main development in extracting voice features and the employment of machine learning models for COVID-19 detection. Related work provides a literature review on respiratory sound analysis and deep learning models for COVID-19 diagnosis. The used datasets, the architecture of the proposed models, and features extracted in this research are explained in Materials and methods. This is followed by a description of the results in Results. In Discussion, we discuss the interpretation of the results and summarise the findings and identifies future directions.

## Background

Voice samples provide a plethora of health-related information [18]. As a result, scientists believe that minor voice signals might reveal underlying medical issues or disease risks. Voice analysis technologies have the potential to be reliable, efficient, affordable, convenient, and simple-to-use techniques for health problem prediction, diagnosis, and monitoring. Various approaches [19, 20] have been used to extract certain voice and acoustic features (referred to as vocal biomarkers) from audio recordings. These features are then analyzed for representative patterns and cues to give insights about an individual's health.

Besides, researchers have investigated several methods to extract sound features for COVID-19 detection. MFCC is a technique to extract audio features that is extensively utilized in different audio recognition applications such as speech emotion identification [21] and pathology voice recognition [22]. Its success stands in the ability to capture in a compact form and in the way human hearing perceives sound [23]. Specifically, MFCC is based on known variations of the human ears' critical bandwidth with frequency. The most important point of speech analysis is that the sounds produced by humans are filtered by the shape of the vocal tract (including the laryngeal cavity, the pharynx, and the oral cavity). The shape of the vocal tract controls how the sound is produced, it reveals in the envelope of the short-time power spectrum, and MFCC is to appropriately capture this envelope.

With the development of deep learning and machine learning, neural networks have played an important role in audio recognition: LSTM [24], SVM [25], CNN [26], artificial neural network (ANN) [27], HuBERT [28] have been widely used for speech analysis. When the HuBERT model is used for speech recognition tasks, it either matches or improves upon the state-of-the-art wav2vec 2.0 performance on all fine-tuning subsets [28]. However, the performance of HuBERT model in downstream applications like vocal pathology detection is not clear. Compared with traditional methods, deep learning, and machine learning models can extract and learn more complex and robust features and make intelligent decisions.

## Related work

Mild or severe changes in human voice can be a sign of a variety of diseases, making vocal biomarkers a noninvasive tool to monitor patients, grade the severity and the stages of diseases, or for drug development [18]. Patients suffering from Parkinson's disease, for example, have a decrease in voice volume and faster speech [29]. As a result, modifications in voice, which are normally undetectable by human ears, can now be investigated by machine learning models. For instance, Wroge et al. [30] implemented a model for the Parkinson's disease diagnosis. Their model can distinguish between Parkinson's disease patients and those in the control group with an accuracy of 85%.

The automatic COVID-19 detection from respiratory sounds has attracted a lot of interest since the outbreak of COVID-19. Hamdi et al. [31] developed a COVID-19 diagnosis system from cough sounds and achieved an area under the ROC curve (AUC) score of 0.9113 with an attention-based hybrid CNN-LSTM model. Kamble et al. [32] used different auditory-based features from breathing, cough, and speech with a Bi-LSTM model to obtain an AUC of 0.866 for COVID-19 detection.

Machine learning and deep learning have been investigated for COVID-19 diagnosis. Lella and Pja [19] proposed to train a multi-channeled deep CNN with features including Data De-noising Auto Encoder (DAE), Gamma-tone Frequency Cepstral Coefficients (GFCC) filter bank, and improved MFCCs (IMFCCs), obtain an accuracy of 95.45% and an F1-score of 96.96%. In [20], the authors studied 76 positive COVID-19 patients and 40 healthy individuals, VGG19 was performed with Mel-spectrograms to distinguish patients with COVID-19. The accuracy for polysyllabic sentence data achieved 81% and for cough data it was 67%. Nassif et al. [16] and Aly et al. [14] both used the Coswara dataset for COVID-19 detection and achieved accuracy of 98.9% and 96% from the CNN models, respectively. Chang et al. [15] proposed a transfer learning framework for the FluSense dataset to detect COVID-19 from cough sounds, the CNN model incorporating embeddings achieved the best validation AUC of 72.38%.

To the best of our knowledge, the works that are fairly comparable with this investigation are [5, 17], since they also use audio recordings from the Cambridge COVID-19 Sound database. In [17], Schuller et al. proposed a technique to identify COVID-19 and achieved an unweighted average recall of 72.1% using a SVM. Han et al. [5] also used the Cambridge COVID-19 Sound database (including breathing, coughs, and voice signals), to obtain an AUC of 0.71 for COVID-19 detection with the VGGish model. Compared with those works, we only use speech recording instead of all breathing, coughs, and speech data. Speech is more natural and tends to have more standardized collection protocols compared to cough sounds, which can vary significantly based on the individual's effort. Collecting breathing sounds requires specialized equipment like a breathing belt, and direct breathing collection from phones is more susceptible to noise compared with speech, complicating the process, and making it impractical for daily monitoring. Additionally, collecting clinical data typically necessitates a visit to a health facility. Therefore, we focus solely on speech signals. We achieve higher AUC performance, the detailed results will be shown in the following sessions.

## Materials and methods

In this section, we will describe the datasets and the models used in this study, and the steps followed by the feature extraction before the analysis. Figure 1 illustrates the model development steps into a pipeline, the MFCC is extracted as inputs for LR, SVM, CNN, and LSTM models, all models were implemented in Python, with detailed descriptions provided in the corresponding sections. From our literature review in Related work, the CNN model with Mel-spectrograms also achieves good performance in image classification and recognition tasks, thus we also train Mel-spectrograms with the CNN model. HuBERT is an end-to-end model which performs various speech and audio processing tasks without the need for handcrafted feature engineering, the raw speech recordings are directly used as inputs for the HuBERT model. The results of these features and models will be discussed in the following sections.

### Dataset description

To develop the classifier models, we use the Cambridge COVID-19 Sound database which is crowd-sourced and collected from a web-based platform, an Android application, and an iOS application. As reported by [10], participants were asked to report their demographics, medical history, and smoking status. In addition, they were required to report their COVID-19 test results, hospitalization status, and symptoms (if any). After a year of data collection, 893 speech samples (308 COVID-19-positive samples) are released. In Figure 2, user characteristics are shown (pnts stands for "prefer not to say"). As shown in Table 1, the demographics and COVID-19 test information are listed.

We develop several classification models to detect COVID-19. For an external validation of the predictive models, we use the Coswara dataset [33]. Details of this dataset are shown in Table S1 in Supplementary material, the sound samples of the Coswara dataset are also collected via worldwide crowdsourcing using a web and mobile application. In our analysis, we focus only on the speech signals of the Cambridge COVID-19 Sound database, thus we also use speech signals in the Coswara dataset (provided as normal and fast speed counting) for model validation.

All the participants in the Cambridge and Coswara datasets were enrolled from online platforms (applications and web). The test results come from self-reports, as the authors claim that although such methodology may result in reduced reliability of the labels, it is significantly more scalable for collecting large amounts of data outside the hospital environment [33, 34].

The COVID test results in the Cambridge dataset are shown in Figure 2c: negative over 14 days, negative in the last 14 days, and negative never, those test results were labeled as negative, and positive over 14 days, positive in the last 14 days, and yes, those test results were labeled as positive. In the Coswara dataset shown in Table S1, the COVID-19 test results such as health, no-resp-illness-exposed, recovered-full were labeled as negative, and positive-mild, positive-moderate, and resp-illness-not-identified were labeled as positive.

Both the Cambridge COVID-19 Sound database and the Coswara dataset are not perfectly balanced in positive and negative cases. All experiments were carried out using 10-fold cross-validation, where one of the folds was used for testing, while the others were used for training. Since the positive and negative cases in datasets are imbalanced, the StratifiedKFold function from scikit-learn library was used to keep all folds the same ratio between positive and negative cases.

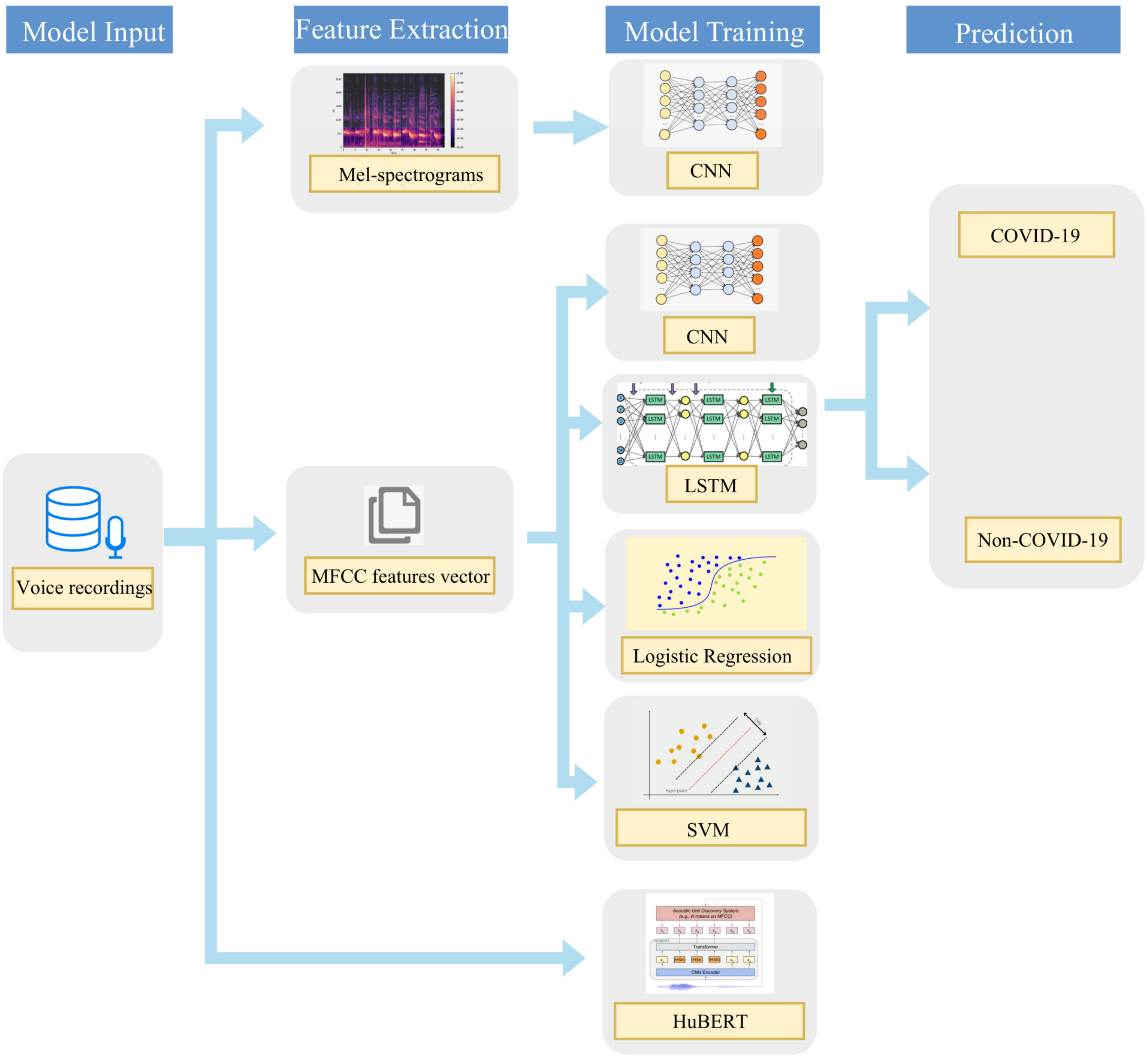

**Figure 1.** The used pipeline for both traditional machine learning classifiers and deep learning classifiers for COVID-19 binary classification

## Models

### Traditional machine learning classifiers

We started with traditional machine learning models as a baseline and progressively built more complex ones. We evaluated LR and SVM as baseline models. The LR and SVM models were implemented using the Scikit-learn library [35]. These models were trained on MFCC features extracted from the audio recordings. All parameters in the LR model were set to their defaults. In the SVM model, the gamma value was set as 0.001 based on fine-tuning with the GridSearchCV library, other parameters in SVM were set to their defaults.

### Convolutional neural network (CNN)

With the development of deep learning, more and more deep learning models are applied to various tasks, such as image recognition, image classification, speech recognition, and machine translation [26]. Due to the outstanding performance of neural networks, CNN has solved several complex challenges in computer vision. Therefore, a CNN model is used to process MFCC images that are extracted from the audio recordings. The CNN model was built based on the TensorFlow 2.13 [36].

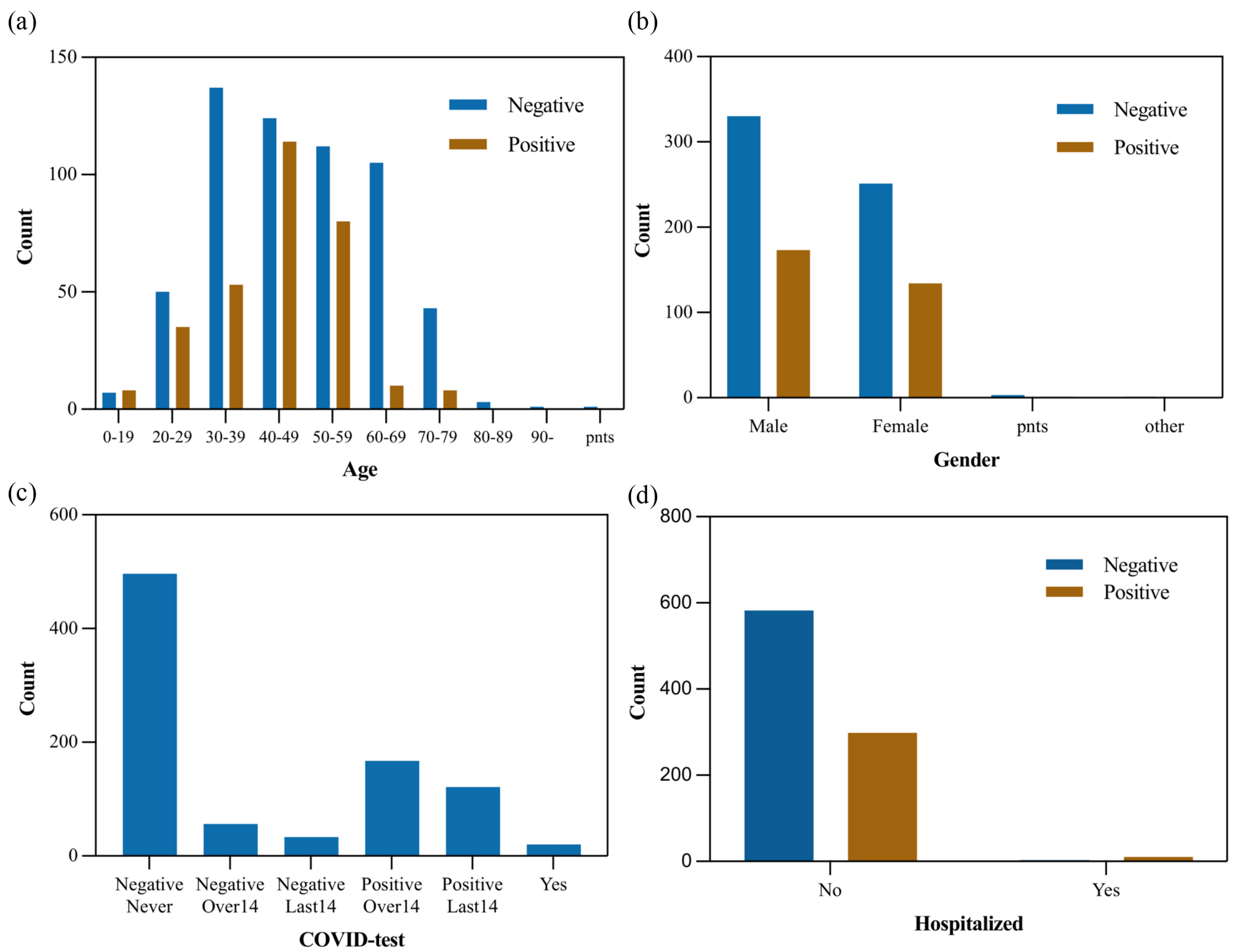

**Figure 2.** Users characteristics (a) age, (b) gender, (c) COVID-19 test results, (d) the number of admissions to hospital

**Table 1.** Patient characteristics in the Cambridge COVID-19 Sound database

| Attribute | Variable | Value | Count | Percentage |
|---|---|---|---|---|
| Demographics | Gender | Female | 385 | 43.1% |
| | | Male | 503 | 56.3% |
| | | Other | 5 | 0.6% |
| | Age | Max age | 90 | |
| | | Min age | 9.5 | |
| | | Avg age | 46 | |
| Hospitalized | | Yes | 13 | 1.5% |
| | | No | 880 | 98.5% |
| COVID-19 test result | | Negative | 585 | 65.5% |
| | | Positive | 308 | 34.5% |
| Recorded time delay | | Positive (recorded in 2 weeks after COVID test) | 141 | 45.8% |
| | | Positive (recorded over 2 weeks after COVID test) | 167 | 54.2% |

We also experimented with Mel-spectrograms as inputs for the CNN model, as the Mel-spectrogram is an effective tool to extract hidden features from audio and visualize them as an image. Moreover, MFCC is less correlated compared with Mel-spectrogram, thus using Mel-spectrogram instead of MFCC achieves better performance for most deep learning models [37]. This results from deep learning models are better equipped to handle correlated features when compared to traditional machine learning models [38].

The CNN model was built with two convolutional layers, each one being followed by a max-pooling layer and dropout layer, finally, a softmax activation function was used in the output dense layer for classification. It is trained by an Adam optimizer, the max epoch and batch size are 100 and 32, respectively.

This shallow CNN model was used as a baseline model, which provides a good starting point to evaluate the performance improvements offered by more advanced models like HuBERT.

### Long short-term memory (LSTM)

LSTM [39] is an advanced variant of recurrent neural network (RNN) that excels in handling sequential data and describing temporal dependencies in the data. The intuition behind this choice is that LSTM allows the neural network to retain (and gradually forget) information about previous time instants taking advantage of the strong temporal dependency that exists between consecutive frames in the speech signals. Our LSTM model was implemented in Python using the Keras library. We extracted MFCC from each audio recording. These features were used to train the LSTM model.

The LSTM model was built with one Bidirectional LSTM layer followed by a dropout layer and two dense layers, finally, a sigmoid activation function was used in the output dense layer for classification. It was trained by an Adam optimizer, the max epoch and batch size are 100 and 32, respectively. The LSTM model was developed with the Scikit-learn library [35].

### Hidden-Unit BERT (HuBERT)

HuBERT is a self-supervised speech model trained to predict clustered features with masked inputs. In this work, we used the base variant pre-trained with 960 h of LibriSpeech [40]. For the pre-training process, a randomly initialized softmax layer is used to replace the projection layer, and then the CTC (connectionist temporal classification) loss is optimized. For more details of the pre-training of HuBERT, please refer to [28]. Though the HuBERT model matches or improves upon the state-of-the-art wav2vec 2.0 performance on fine-tuning speech recognition task [28], its performance on vocal pathology tasks is not clear. We used the HuBERT model in this work to explore its effectiveness in vocal pathology tasks.

The base-HuBERT model used in this work has 12 layers of transformers and contains 90M parameters. The HuBERT model implemented in this work is hosted on the Hugging Face model hub. The entire HuBERT is fine-tuned with the Cambridge COVID-19 Sound database, and the results will be compared with other machine learning models.

### Feature extraction

Vocal features were extracted from audio recordings and used in our models. The extracted features for speech signal optimum representation are the MFCC, Mel-spectrograms and CNN Encoder features.

MFCC is a fundamental feature that is utilized in speaker and emotion recognition and disease detection by the advanced representation of human auditory perception it provides [21, 22]. MFCC was extracted from each voice recording in frames by using the default frame length of 2,048 samples, and hop length of 512 samples, the first 40 cepstral coefficients were computed in each frame. For the CNN model, the MFCC images were used as inputs, but for LR, SVM, and LSTM models, MFCC vectors were obtained by computing the mean from the frame-wise MFCC, because images as inputs are unadmissible for these models.

Mel-spectrogram [41] contains a short-time Fourier transform (STFT) for each frame of the spectrum (energy/amplitude spectrum), from the linear frequency scale to the logarithmic Mel-scale, the filter bank will then determine the eigenvector. Eigenvalues can be roughly expressed as the distribution of signal energy on the Mel-scale frequency. Mel-spectrogram was extracted from each voice recording in frames by using the default frame length of 2,048 samples, and hop length of 512 samples, Hanning window was used to control the spectral leakage.

HuBERT model is an end-to-end model, the CNN Encoder in HuBERT was composed of seven 512-channel layers with strides [5,2,2,2,2,2,2] and kernel widths [10,3,3,3,3,2,2]. The CNN Encoder generated a feature sequence at a 20 ms framerate and then randomly masked as inputs.

## Results

### Performance measurements

To better evaluate the performance of the model, we listed several indicators used to evaluate the model. Among the indicators, AUC and accuracy were considered as the most important performance indicators. AUC is particularly valued because it considers all possible classification thresholds, making it insensitive to the choice of threshold values. Moreover, AUC is a more robust indicator due to the fact that it is less sensitive to imbalanced datasets. Accuracy was also considered an important indicator since it represents the percentage of correctly classified cases, making it easy for stakeholders, healthcare professionals, and the general public to understand the model's performance.

### Experimental results with different models

We trained the models, and the results are summarized in Table 2, the detailed results for each folder are shown in Table S2 to Table S7 in Supplementary material. The accuracy of the CNN model with MFCC images is 59%. For positive cases, 38% of them are identified accurately as positive. 69% of those who tested negative are expected to be COVID-19 free, and the positive predictive value (PPV) and negative predictive value (NPV) of the CNN model are 33% and 72%. Figure 3 shows the ROC curve for the CNN classifier, the curve shows performance with 0.54 AUC. We also used Mel-spectrograms as inputs for the CNN model, the accuracy and AUC are higher than the CNN model with MFCC, which are 78% and 0.84. It indicates when Mel-spectrograms are used as inputs, the CNN model has better performance for COVID-19 detection.

**Table 2.** A summary of parameters and performances of the used models

| Model | Parameters | Accuracy | Sensitivity | Specificity | PPV | NPV |
|---|---|---|---|---|---|---|
| LR | Input = MFCC vector | 0.71 ± 0.04 | 0.62 ± 0.11 | 0.73 ± 0.03 | 0.39 ± 0.12 | 0.87 ± 0.05 |
| SVM | Input = MFCC vector, kernel = rbf, C = 1, gamma = 0.001 | 0.81 ± 0.04 | 0.87 ± 0.01 | 0.80 ± 0.03 | 0.54 ± 0.08 | 0.96 ± 0.03 |
| CNN | Input = MFCC images, input shape = (150,150,3), loss = binary crossentropy, optimizer = adam, activation = softmax | 0.59 ± 0.11 | 0.38 ± 0.32 | 0.69 ± 0.31 | 0.33 ± 0.20 | 0.72 ± 0.11 |
| LSTM | Input = MFCC vector, loss = mean absolute error, optimizer = adam, activation = sigmoid | 0.81 ± 0.03 | 0.63 ± 0.06 | 0.90 ± 0.04 | 0.77 ± 0.08 | 0.83 ± 0.03 |
| CNN | Input = Mel-spectrogram images, input shape = (150,150,3), loss = binary crossentropy, optimizer = adam, activation = softmax | 0.78 ± 0.03 | 0.65 ± 0.12 | 0.85 ± 0.04 | 0.70 ± 0.04 | 0.82 ± 0.04 |
| HuBERT | Input = Encoder features | 0.86 ± 0.03 | 0.80 ± 0.09 | 0.89 ± 0.07 | 0.82 ± 0.08 | 0.90 ± 0.04 |

We tested the performance of the LSTM model, and we took the same strategy as the MFCC features extracted from the audio recordings. According to Table 2, the proposed LSTM model has the highest accuracy of 81%, with 63% sensitivity and 90% specificity. LSTM model produces PPV and NPV as 77% and 83%. Figure 3 shows the LSTM model has an AUC of 0.80.

We also trained two traditional classifiers to predict COVID-19 based on MFCC features. According to Table 2, after we removed some delayed positive cases, the LR model has an accuracy of 71%, correctly recognizes 62% of positive cases, and the model accurately classifies 73% of negative samples. Figure 3 shows the LR model with a 0.72 AUC. Table 2 shows for the SVM model, the accuracy is 81%, with 87% sensitivity and 80% specificity.

The best performance was achieved by HuBERT model with 86% accuracy and 0.93 of AUC. Those results show when the MFCC features were used as inputs, the SVM model has the highest accuracy and AUC among all models. Overall, the HuBERT model achieves the highest AUC and accuracy of 0.93 and 0.86 compared with other models.

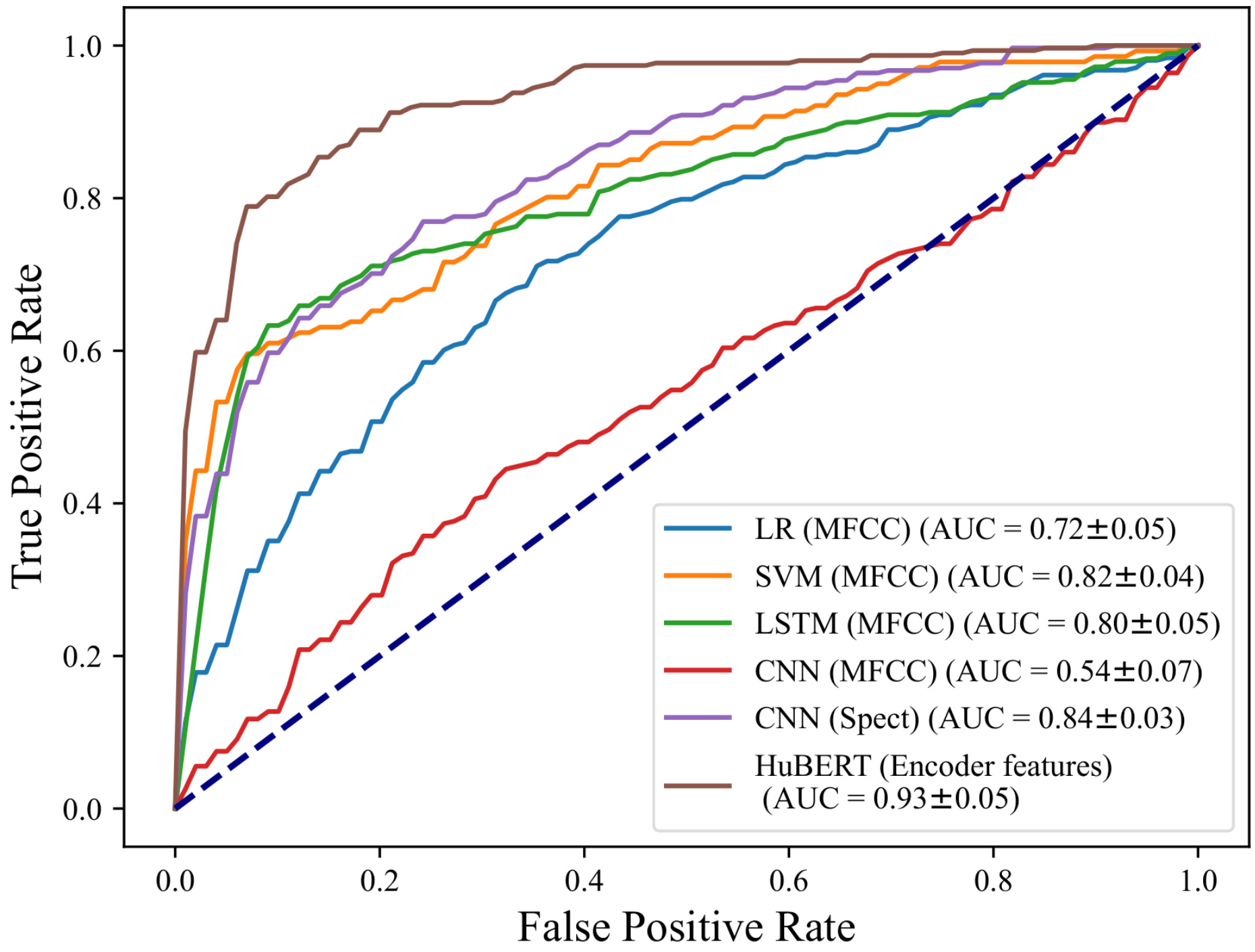

**Figure 3.** ROC curve for models

### Coswara dataset validation

From previous sections, the HuBERT model performs better since it has the highest AUC and accuracy, but LSTM and CNN also show good performance with high accuracy and AUC, so we validate all of those three models. We retrained the HuBERT, LSTM model and CNN model with the Coswara dataset to verify the performance of those models. Again, MFCC features and MEL-spectrograms of the Coswara dataset were extracted and fed into LSTM and CNN models.

There are 3,898 negative cases and 1,350 positive cases in the Coswara dataset. As Table 3 shows, when we applied our proposed LSTM model to the Coswara dataset, we got an accuracy of 75%, sensitivity and specificity are 29% and 93%. Figure 4 shows LSTM with an AUC of 0.66. The CNN model was also used for validation and Mel-spectrograms were used as inputs, we achieved an accuracy of 71% and an AUC of 0.71. The HuBERT model achieves the highest performance in all indicators including accuracy of 0.82 and AUC of 0.83, which shows the HuBERT model also has the best performance in COVID detection for the external dataset.

**Table 3.** A summary of the performances and parameters for Coswara dataset validation

| Model | Parameters | Accuracy | Sensitivity | Specificity | PPV | NPV |
|---|---|---|---|---|---|---|
| LSTM | Input = MFCC vector, loss = mean absolute error, optimizer = adam, activation = sigmoid | 0.75 ± 0.06 | 0.29 ± 0.03 | 0.93 ± 0.02 | 0.61 ± 0.07 | 0.79 ± 0.01 |
| CNN | Input = Mel-spectrogram images, input shape = (150,150,3), loss = binary crossentropy, optimizer = adam, activation = softmax | 0.71 ± 0.03 | 0.44 ± 0.04 | 0.84 ± 0.03 | 0.49 ± 0.05 | 0.81 ± 0.01 |
| HuBERT | Input = Encoder features | 0.82 ± 0.03 | 0.50 ± 0.09 | 0.93 ± 0.05 | 0.73 ± 0.10 | 0.84 ± 0.02 |

### Distinguishing COVID-19 from cold symptoms

To make sure our models are performing well in classifying COVID-19 instead of other cold symptoms, we used the proposed models on all cases that tested COVID-19 positive, and on those cases that tested as COVID-19 negative but reported at least one cold symptom, such as dry cough, wet cough, fever, or sore throat. There are 308 positive cases and 216 negative with cold symptom cases left in this experiment.

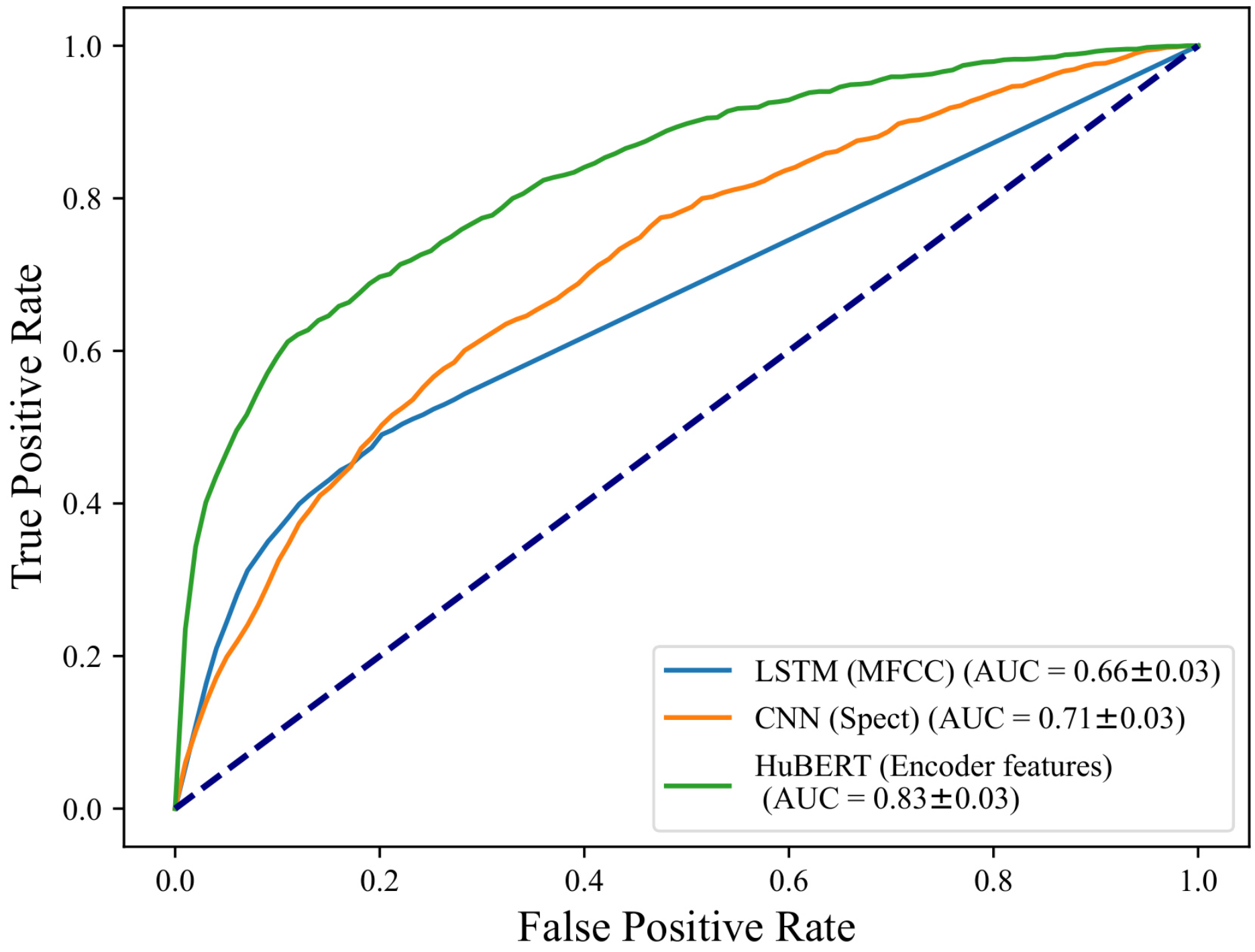

**Figure 4.** ROC curve for Coswara dataset validation

The results in Figure 5 show an AUC of 0.78 for the LSTM model and an AUC of 0.85 for the CNN model with Mel-spectrograms as inputs, the highest AUC is achieved by HuBERT model with 0.90, showing that the negative cases with cold symptoms can be distinguished from COVID-19 positive quite well. In other words, our models can distinguish COVID-19 cases in voices without misclassifying subjects with cold symptoms.

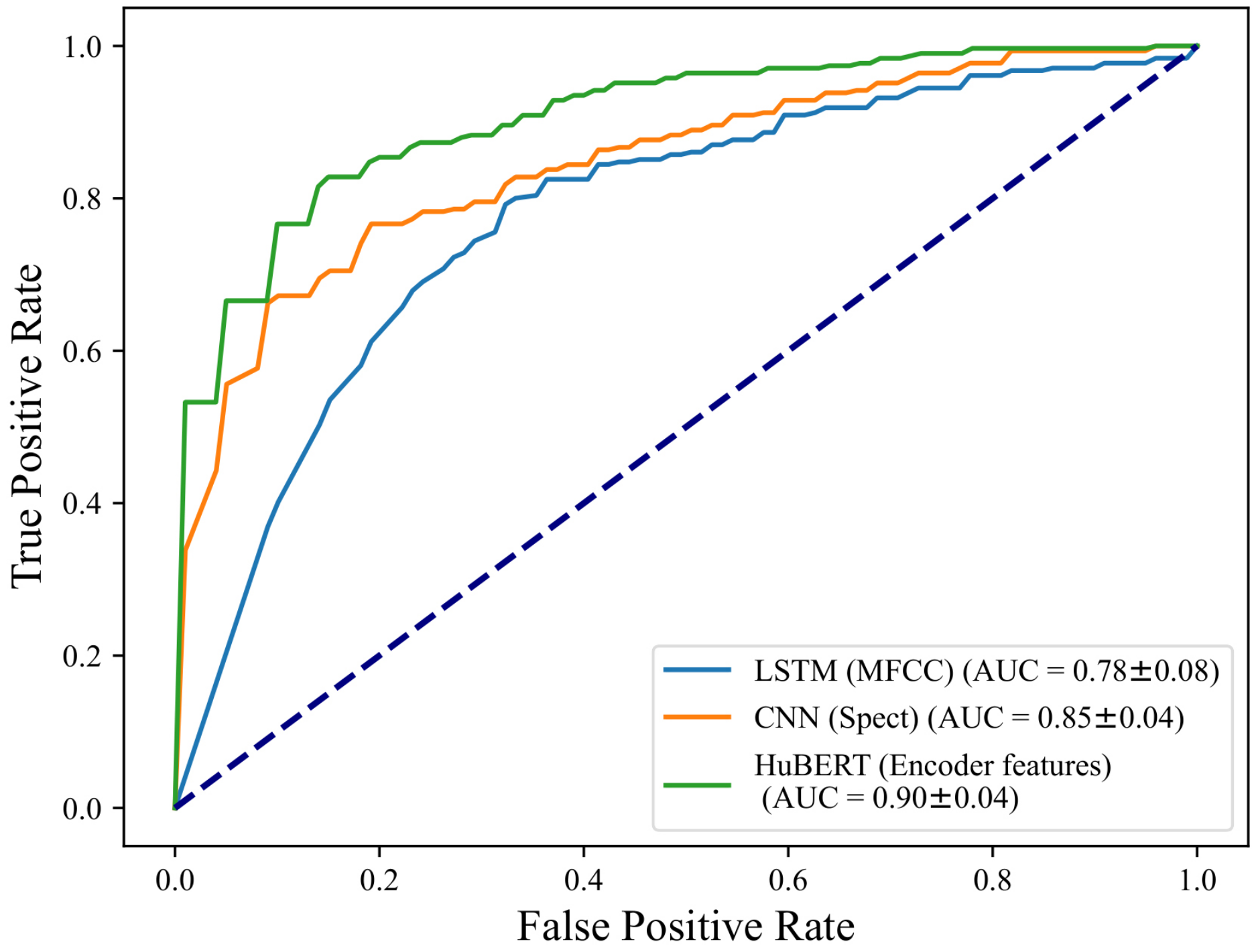

**Figure 5.** ROC curve for distinguishing COVID-19 from cold symptoms

## Discussion

In this work, we extracted Mel-spectrograms, MFCC, and CNN Encoder features for deep learning models and traditional machine learning models. We demonstrated that the HuBERT model can best distinguish

COVID-19 patients. We concluded this based on the HuBERT model achieved the highest AUC and accuracy of 0.93 and 0.86. The HuBERT model could be used as a complementary method for COVID-19 screening. The best performance from the HuBERT model can be attributed to several factors. Firstly, compared with other models, HuBERT uses a more deep network architecture that experts in learning intricate patterns and features from raw audio data. It combines CNN Encoder and transformer layers, enabling it to capture both local acoustic cues and long-range dependencies in voice. Secondly, the Hubert model benefits from pre-training on a large-scale dataset (960 h of LibriSpeech), and the diversity of the pre-training dataset allows it to acquire general knowledge from various audio sources. This transfer learning equips HuBERT with a strong foundation to identify subtle voice changes associated with COVID-19. Moreover, HuBERT is end-to-end, there is no need for handcrafted feature engineering, and enables it to learn directly from voice recordings without the loss of information. This is especially important for tasks like COVID-19 detection, where the detailed vocal bio-markers may not be well-defined.

To verify the performance of our proposed models, we trained the proposed models with a new dataset (Coswara), the HuBERT model achieved the best performance again, despite the two datasets being collected with different processes. One clear difference between the two datasets was that the recording in Coswara involved counting to twenty, rather than repeating 3 times a phrase. This shows that the HuBERT model has good generalization. We also validated our models with positive COVID-19 cases and negative cases but reported at least one cold symptom, the 0.90 AUC shows the HuBERT model is truly detecting COVID-19 instead of cold symptoms.

We addressed potential biases due to demographic and environmental factors through several strategies. First, we utilized the Cambridge COVID-19 Sound dataset, which includes a diverse global population with varying demographics such as age, gender, hospitalization history, and medical history, as detailed in Table 1. The crowd-sourced nature of this dataset, collected from different platforms, ensured a variety of recording environments ranging from quiet rooms to noisy backgrounds. This diversity enhances the robustness of our model, allowing it to generalize effectively across different populations. Additionally, we validated our model using the Coswara dataset, which features a distinct demographic profile, particularly in terms of language and content in speech, as shown in Table S1. These demographic differences further reinforce the robustness of our model. Furthermore, we employed cross-validation techniques, including the StratifiedKFold function, to assess the model's performance across different data subsets. Those approaches provide a comprehensive and generalizable performance measure, ensuring our model's reliability and robustness across diverse demographic and environmental conditions.

Moreover, the nature of the data itself did not allow us to consider additional variables that may influence the test results. For example, in this work, we could not control for variables such as overstrain of the vocal cords, laryngeal dystonia, or gastroesophageal reflux. Nevertheless, we expect our model to be robust to small variations in an individual's voice which may naturally change over time. This is because our model works with extracted voice features such as MFCC which have been proven to be robust to small changes of minor daily and environmental variations in voice [42].

Generally, the above findings show promising results for COVID-19 detection from voice recordings, and this study provides a low-cost, non-invasive, and ubiquitous detection method, especially for underdeveloped and low-income regions. If we compare the sensitivity and specificity of the HuBERT model to the well-known RAT, the overall sensitivity of RAT for COVID-19 detection is 65% [13], and our proposed model has a higher sensitivity (80%). High sensitivity implies that few cases are misclassified as negative, this can avoid COVID-19 spreading. On the other hand, the HuBERT has a high specificity of 89%, however, it is lower than the specificity of RAT (99%) which means that our test might miss-classify some negative patients as positive. When we compare these results, we also need to be cautious as, in the case of the RAT performance indicators [13], the results are more robust due to the larger set of data available for analysis, therefore, more voice data would be needed to confidently compare our model to the RAT test. For breath analyzers, the accuracy for COVID-19 detection is 80–81.5% [43], which is comparable with 82% of our HuBERT model. However, in terms of ease of use, it is difficult to collect breath and avoid contamination from the container, what's more, a portable gas chromatograph-mass spectrometer is needed, which also increases the cost of diagnosis. Overall, our test would be non-invasive to users and could be virtually provided at little or no cost for complementary methods for COVID-19 screening in the post-COVID era. This test can be used for large-scale surveillance, helping to identify and track the spread of COVID-19 in communities. By further integrating the model into mobile apps and online platforms in the future, it may be possible to increase public awareness and encourage self-monitoring and timely medical consultation.

Privacy preserving is important in voice data collection and analysis. Both the Cambridge and Coswara datasets used in this work were collected following ethical approval from their respective Ethics Committee, ensuring compliance with ethical guidelines and standards. All the data collected was anonymized and excluded any participant identity information. For the analysis, the machine models in this work do not require the storage of raw voice data or any sensitive information such as demographic or clinical data from the patients. Instead, the analysis involves transforming voice data into voice features. This transformation abstracts the data to a level where the original voice cannot be easily reconstructed, thus enhancing privacy. This work focuses on developing a robust and accurate model for COVID-19 screening, future implementation of this work can be integrated with additional cloud privacy-preserving techniques to further ensure the safety of the analysis.

In conclusion, non-intrusive and easy-to-collect voice features may provide valuable information for COVID-19 screening in the post-COVID era. Using voice features extracted from audio recordings we defined a predictive model that can classify COVID-19. This work suggests that the best performance is obtained with the HuBERT model with an AUC and sensitivity of respectively 0.93 and 0.80, this may result from the HuBERT's ability to learn high-level representations directly from the data can lead to improved performance. However, compared to traditional machine learning models, the HuBERT model may suffer from limited interpretability. Therefore, it is important to balance both performance and interpretability in practical use. These results are competitive with recent COVID-19 cough/breathing identification studies [5, 17], references [5] obtained an AUC of 0.71, which is lower than the 0.93 AUC achieved by our Hubert model. Reference [17] achieved the recall of 72.1%, which is lower than the 80% obtained by the Hubert model, and indicate voice recording can be a timely, low-cost, and safe screening tool for COVID-19.

In future works, we plan to further explore the interpretability of how the AI models perform the classification, we will study which are the relevant voice features that characterize COVID-19 patients. The SHAP (SHapley Additive exPlanations) and LIME (local interpretable model-agnostic explanations) can be used to further provide insights into the features in our models including HuBERT. Furthermore, we will perform interdisciplinary studies with respiratory physicians to include their domain knowledge in feature designing and performance valuation, and in linking the extracted features associated with COVID-19 to other respiratory conditions.

## Abbreviations

AUC: area under the ROC curve

CNN: convolutional neural network

COVID-19: coronavirus disease 2019

HuBERT: Hidden-Unit BERT

LR: logistic regression

LSTM: long short-term memory

MFCC: Mel-frequency cepstral coefficients

NPV: negative predictive value

PPV: positive predictive value

RAT: rapid antigen test

RT-PCR: real-time reverse transcription polymerase chain reaction

SARS-CoV-2: severe acute respiratory syndrome coronavirus 2

SVM: support vector machine

## Supplementary materials

The supplementary material for this article is available at: https://www.explorationpub.com/uploads/Article/file/101122_sup_1.pdf.

## Declaration

### Acknowledgments

The authors would like to thank Cambridge University for sharing the data collected via the COVID-19 app, and the Indian Institute of Science Bangalore for opening the Coswara dataset which supported our model validation.

### Author contributions

YY: Methodology, Software, Formal analysis, Writing—original draft, Investigation. WA: Conceptualization, Methodology, Software, Writing—original draft. SOS: Supervision, Writing—review & editing. VU: Methodology, Supervision, Writing—review & editing.

### Conflicts of interest

The authors declare that they have no conflicts of interest.

### Ethical approval

This study was approved by the Ethical Review Committee Inner City Faculties (ERCIC), the reference number is ERCIC_528_31_01_2024.

### Consent to participate

Informed consent was obtained by the data sources for the study.

### Consent to publication

Not applicable.

### Availability of data and materials

The Cambridge dataset is accessed by a Data-Sharing Agreement with Cambridge University. The Coswara dataset is public, and can be accessed online: https://github.com/iiscleap/Coswara-Data.

### Funding

The work was supported by NWO Aspasia grant [91716421]. The funders had no role in study design, data collection and analysis, decision to publish, or preparation of the manuscript.

## References

1. COVID-19 coronavirus pandemic [Internet]. Worldometers.info; c2024 [cited 2024 Feb 1]. Available from: https://www.worldometers.info/coronavirus/
2. Lai CC, Shih TP, Ko WC, Tang HJ, Hsueh PR. Severe acute respiratory syndrome coronavirus 2 (SARS-CoV-2) and coronavirus disease-2019 (COVID-19): The epidemic and the challenges. Int J Antimicrob Agents. 2020;55:105924. [DOI] [PubMed] [PMC]

3. Science brief: SARS-CoV-2 and surface (fomite) transmission for indoor community environments [Internet]. [Cited 2024 Feb 1]. Available from: https://stacks.cdc.gov/view/cdc/104762
4. Ningthoujam R. COVID 19 can spread through breathing, talking, study estimates. Curr Med Res Pract. 2020;10:132–3. [DOI] [PubMed] [PMC]
5. Han J, Xia T, Spathis D, Bondareva E, Brown C, Chauhan J, et al. Sounds of COVID-19: exploring realistic performance of audio-based digital testing. NPJ Digit Med. 2022;5:16. [DOI] [PubMed] [PMC]
6. Stasak B, Huang Z, Razavi S, Joachim D, Epps J. Automatic Detection of COVID-19 Based on Short-Duration Acoustic Smartphone Speech Analysis. J Healthc Inform Res. 2021;5:201–17. [DOI] [PubMed] [PMC]
7. Hassan A, Shahin I, Alsabek MB. COVID-19 detection system using recurrent neural networks. In: The 2020 International Conference on Communications, Computing, Cybersecurity, and Informatics; 2020 Nov 3-5; Sharjah, United Arab Emirates. 2020. [DOI]
8. Mehrabadi MA, Aqajari SAH, Azimi I, Downs CA, Dutt N, Rahmani AM. Detection of COVID-19 Using Heart Rate and Blood Pressure: Lessons Learned from Patients with ARDS. Annu Int Conf IEEE Eng Med Biol Soc. 2021;2021:2140–3. [DOI] [PubMed] [PMC]
9. Liang JS, Wang K. Vibration feature extraction using audio spectrum analyzer based machine learning. In: 2017 International conference on information, Communication and Engineering (ICICE); 2017 Nov 17-20; Xiamen, China. IEEE; 2017. pp. 381–4. [DOI]
10. Brown C, Chauhan J, Grammenos A, Han J, Hasthanasombat A, Spathis D, et al. Exploring automatic diagnosis of COVID-19 from crowdsourced respiratory sound data. ArXiv [Preprint]. 2021 [cited 2024 Feb 1]. Available from: https://arxiv.org/abs/2006.05919
11. Vahedian-Azimi A, Keramatfar A, Asiaee M, Atashi SS, Nourbakhsh M. Do you have COVID-19? An artificial intelligence-based screening tool for COVID-19 using acoustic parameters. J Acoust Soc Am. 2021;150:1945–53. [DOI]
12. Despotovic V, Ismael M, Cornil M, Call RM, Fagherazzi G. Detection of COVID-19 from voice, cough and breathing patterns: Dataset and preliminary results. Comput Biol Med. 2021;138:104944. [DOI] [PubMed] [PMC]
13. Arshadi M, Fardsanei F, Deihim B, Farshadzadeh Z, Nikkhahi F, Khalili F, et al. Diagnostic Accuracy of Rapid Antigen Tests for COVID-19 Detection: A Systematic Review With Meta-analysis. Front Med (Lausanne). 2022;9:870738. [DOI] [PubMed] [PMC]
14. Aly M, Rahouma KH, Ramzy SM. Pay attention to the speech: COVID-19 diagnosis using machine learning and crowdsourced respiratory and speech recordings. Alexandria Eng J. 2022;61:3487–500. [DOI]
15. Chang Y, Jing X, Ren Z, Schuller BW. CovNet: A Transfer Learning Framework for Automatic COVID-19 Detection From Crowd-Sourced Cough Sounds. Front Digit Health. 2022;3:799067. [DOI] [PubMed] [PMC]
16. Nassif AB, Shahin I, Bader M, Hassan A, Werghi N. COVID-19 detection systems using deep-learning algorithms based on speech and image data. Mathematics. 2022;10:564. [DOI]
17. Schuller BW, Batliner A, Bergler C, Mascolo C, Han J, Lefter I, et al. The INTERSPEECH 2021 computational paralinguistics challenge: COVID-19 cough, COVID-19 speech, escalation & primates. ArXiv [Preprint]. 2021 [cited 2024 Feb 1]. Available from: https://arxiv.org/pdf/2102.13468
18. Fagherazzi G, Fischer A, Ismael M, Despotovic V. Voice for Health: The Use of Vocal Biomarkers from Research to Clinical Practice. Digit Biomark. 2021;5:78–88. [DOI] [PubMed] [PMC]
19. Lella KK, Pja A. Automatic diagnosis of COVID-19 disease using deep convolutional neural network with multi-feature channel from respiratory sound data: cough, voice, and breath. Alexandria Eng J. 2022;61:1319–34. [DOI]
20. Suppakitjanusant P, Sungkanuparph S, Wongsinin T, Virapongsiri S, Kasemkosin N, Chailurkit L, et al. Identifying individuals with recent COVID-19 through voice classification using deep learning. Sci Rep. 2021;11:19149. [DOI] [PubMed] [PMC]

21. Bromuri S, Henkel AP, Iren D, Urovi V. Using AI to predict service agent stress from emotion patterns in service interactions. J Ser Manag. 2021;32:581–611. [DOI]
22. Verma V, Benjwal A, Chhabra A, Singh SK, Kumar S, Gupta BB, et al. A novel hybrid model integrating MFCC and acoustic parameters for voice disorder detection. Sci Rep. 2023;13:22719. [DOI] [PubMed] [PMC]
23. Logan B. Mel frequency cepstral coefficients for music modeling. Proc of Ismir. 2000.
24. Hochreiter S, Schmidhuber J. Long short-term memory. Neu Comp. 1997;9:1735–80. [DOI]
25. Cortes C, Vapnik V. Support-vector networks. Mach Learn. 1995;20:273–97. [DOI]
26. O'Shea K, Nash R. An introduction to convolutional neural networks. arXiv [Preprint]. 2015 [cited 2024 Feb 1]. Available from: https://arxiv.org/abs/1511.08458
27. McCulloch WS, Pitts W. A logical calculus of the ideas immanent in nervous activity. Bull Math Biophys. 1943;5:115–33. [DOI]
28. Hsu WN, Bolte B, Tsai YHH, Lakhotia K, Salakhutdinov R, Mohamed A. Hubert: Self-supervised speech representation learning by masked prediction of hidden units. arXiv [Preprint]. 2021 [cited 2024 Feb 1]. Available from: https://doi.org/10.48550/arXiv.2106.07447
29. Solana-Lavalle G, Rosas-Romero R. Analysis of voice as an assisting tool for detection of Parkinson's disease and its subsequent clinical interpretation. Bio Sig Proc Cont. 2021;66:102415. [DOI]
30. Wroge TJ, Özkanca Y, Demiroglu C, Si D, Atkins DC, Ghomi RH. Parkinson's disease diagnosis using machine learning and voice. In: 2018 IEEE signal processing in medicine and biology symposium (SPMB); 2018 Dec 01; Philadelphia, PA, USA. IEEE; 2018. pp. 1–7. [DOI]
31. Hamdi S, Oussalah M, Moussaoui A, Saidi M. Attention-based hybrid CNN-LSTM and spectral data augmentation for COVID-19 diagnosis from cough sound. J Intell Inf Syst. 2022;59:367–89. [DOI] [PubMed] [PMC]
32. Kamble MR, Patino J, Zuluaga MA, Todisco M. Exploring auditory acoustic features for the diagnosis of covid-19. arXiv [Preprint]. 2022 [cited 2024 Feb 1]. Available from: https://arxiv.org/pdf/2201.09110
33. Sharma N, Krishnan P, Kumar R, Ramoji S, Chetupalli SR, Ghosh PK, et al. Coswara--a database of breathing, cough, and voice sounds for COVID-19 diagnosis. arXiv [Preprint]. 2020 [cited 2024 Feb 1]. Available from: https://arxiv.org/pdf/2005.10548v2
34. Xia T, Spathis D, Brown C, Chauhan J, Grammenos A, Han J, et al. COVID-19 sounds: a large-scale audio dataset for digital respiratory screening. In: 35th Conference on Neural Information Processing Systems (NeurIPS 2021) Track on Datasets and Benchmarks; 2021.
35. Kramer O. Scikit-Learn. In: Machine Learning for Evolution Strategies. Springer, Cham; 2016. pp. 45–53. [DOI]
36. Singh P, Manure A. Introduction to tensorflow 2.0. In: Learn TensorFlow 20. Apress, Berkeley, CA; 2020. pp. 1–24.
37. Huzaifah M. Comparison of time-frequency representations for environmental sound classification using convolutional neural networks. arXiv [Preprint]. 2017 [cited 2024 Feb 1]. Available from: https://arxiv.org/abs/1706.07156
38. Nallanthighal VS. Respiratory health sensing from speech [dissertation]. Amsterdam: LOT; 2022. [DOI]
39. Gers FA, Schmidhuber J, Cummins F. Learning to forget: continual prediction with LSTM. Neural Comput. 2000;12:2451–71. [DOI] [PubMed]
40. Panayotov V, Chen G, Povey D, Khudanpur S. Librispeech: an asr corpus based on public domain audio books. In: 2015 IEEE International Conference on Acoustics, Speech and Signal Processing (ICASSP); 2015 Apr 19-24; South Brisbane, QLD, Australia. IEEE; 2015. pp. 5206–10. [DOI]
41. Aly M, Alotaibi NS. A novel deep learning model to detect COVID-19 based on wavelet features extracted from Mel-scale spectrogram of patients' cough and breathing sounds. Inf Med Unl. 2022;32:101049. [DOI]

42. Kamarulafizam I, Salleh SH, Najeb J, Ariff AK, Chowdhury A. Heart sound analysis using MFCC and time frequency distribution. In: Magjarevic R, Nagel JH, editors. World Congress on Medical Physics and Biomedical Engineering 2006; 2006 Aug 27-Sep 1. Springer, Berlin, Heidelberg; 2006. pp. 946–9. [DOI]
43. Xue C, Xu X, Liu Z, Zhang Y, Xu Y, Niu J, et al. Intelligent COVID-19 screening platform based on breath analysis. J Breath Res. 2022;17:016005. [DOI] [PubMed]